\documentclass{article}[12 pt]
\usepackage{latexsym}

\usepackage{fullpage}

\newtheorem{theorem}{Theorem} 
\newtheorem{lemma}[theorem]{Lemma}
 
\newenvironment{proof}{{\bf Proof: }}{$\Box$}
\newenvironment{remark}{\textbf{Remark: }}{}
\newtheorem{definition}{Definition}

\begin{document}
\title{Multiparty computation unconditionally secure against ${\cal
Q}^2$ adversary structures}
      
\author{Adam Smith%
  \thanks{School of Computer Science, McGill University, Montr\'eal
    (Qu\'ebec), Canada, {\sf asmith@cs.mcgill.ca}} \and
        Anton Stiglic%
        \thanks{D\'epartement d'Informatique et R.O., Universit\'e de
          Montr\'eal, Montr\'eal (Qu\'ebec), Canada, {\sf
            stiglic@iro.umontreal.ca}}}

      \maketitle

\begin{abstract}
  We present here a generalization of the work done by Rabin and
  Ben-Or in \cite{RB89}. We give a protocol for multiparty computation
  which tolerates any ${\cal Q}^2$ active adversary structure based on
  the existence of a broadcast channel, secure communication between
  each pair of participants, and a monotone span program with
  multiplication tolerating the structure. The secrecy achieved is
  unconditional although we allow an exponentially small probability
  of error.  This is possible due to a protocol for computing the
  product of two values already shared by means of a homomorphic
  commitment scheme which appeared originally in \cite{ChaumEG1987}.
\end{abstract}

\textbf{Keywords:} Multiparty computation, general adversary
structures, span programs, verifiable secret sharing.

\section{Introduction}
\label{sec:intro}

\subsection{Multiparty computation}
\label{sec:MPCintro}

Multiparty computation (MPC) is a cryptographic task that allows a
network of participants to emulate any trusted party protocol. Each
player $P_i$ starts with a private input $x_i$. The players run a
protocol to compute some function $g(x_1,\ldots,x_n)$. The result of
this function can then be revealed publicly or privately to some
particular player. The protocol is deemed secure if cheating parties
can obtain no more information from running the protocol than they
would in the trusted party scenario (in which each player gives $x_i$
to some external trusted party who then computes $g$ and sends the
result to all the relevant players). Goldreich, Micali and Wigderson
proved that to accomplish MPC it is sufficient to always have the
value of $g$ revealed publicly and to assume that $g$ is given by an
arithmetic circuit (i.e. addition and multiplication gates) from $K^n$
to $K$ where $K$ is some finite field.

The first general solution to this problem was given in \cite{GMW87}.
They present a protocol for MPC which is secure under the assumptions
that trapdoor one-way permutation exists, that the participants are 
restricted to probabilistic polynomial time (computationally bounded) 
and that the number of cheating parties is
bounded above by $t$ where $t<n/2$. In the situation where the
participants can only cheat passively (i.e. by eavesdropping) they can
remove the last assumption. In \cite{BGW88} and \cite{CCD88}, the
assumption of computational boundedness is removed and replaced by the
assumption that each pair of players is connected by an authenticated
secure channel. In this (non-computational) model they prove that MPC
is possible with active adversaries if and only if $t<n/3$ and with
passive adveraries if and only if $t<n/2$. 

These results were extended in \cite{RB89} to the scenario in which a
reliable broadcast channel is also available.  In that case active and
passive cheaters can be tolerated if and only if $t<n/2$. However, to
attain these bounds an exponentially small probability of error was
introduced.

The result of \cite{RB89} was first extended to more general adversary
structures by Hirt and Maurer in \cite{HM97}. However, maintaining an
exponentially small probability of error entailed a superpolynomial
loss of efficiency. \newpage We present a more efficient version of an
extension of the \cite{RB89} protocol using monotone span programs,
following the ideas of \cite{CDM98} \footnote{Results similar to those
in this article have been found independently in \cite{CCDHR98} }.  The relevant
definitions as well as a precise statement of our results are
presented in the remainder of this section.

\subsection{Adversary structures and monotone functions}
\label{sec:struct}

Given a set of players $P$, an adversary structure ${\cal A}$ over $P$
is a set of subsets of players which is downward-closed under
inclusion:
$$(B \in {\cal A} \textrm{ and } B' \subseteq B) \Longrightarrow B'
\in {\cal A}.$$
Normally such a structure is used to represent the
collection of all coalitions of players which a given protocol can
tolerate without losing security: as long as the set of cheating
players is in ${\cal A}$, the cheaters cannot breach the security of
the protocol.

Classically, protocols such as those of \cite{RB89} have tolerated
\emph{threshold structures}, which are of the form ${\cal A} = \{ B
\subseteq P : |B| \leq t \}$ for some $t$. However, \cite{HM97}
extends several of these results to more general structures, using the
following definition:

\begin{definition}
  An adversary structure ${\cal A}$ over $P$ is said to be ${\bf {\cal
    Q}^k}$ if no $k$ sets in ${\cal A}$ add up to the whole set $P$,
  that is
  $$\not \exists B_1,B_2,\ldots,B_k \in {\cal A} : \quad B_1 \cup B_2
  \cup \cdots \cup B_k = P.$$
\end{definition}

Hirt and Maurer (\cite{HM97}) extended the results of
\cite{BGW88,RB89} (see section \ref{sec:MPCintro}) which held for $t <
n/3$ and $t < n/2$ to ${\cal Q}^3$ and ${\cal Q}^2$ structures
respectively.

\subsubsection{Monotone functions}
\label{sec:mf}

\begin{definition}
  For a partial order $\leq$ on sets $A$ and $B$, we say that a function
  $f: A \rightarrow B$ is {\bf monotone} if for $x,y \in A$ we have
  $$x \leq y \Longrightarrow f(x) \leq f(y)$$
\end{definition}

We can define a partial order on $\{ 0,1 \} ^n$ by the rule ``${\bf x}
\leq {\bf y}$ iff each coordinate of ${\bf x}$ is smaller than the
corresponding coordinate of ${\bf y}$.''  Then a function $f: \{ 0,1 \}
^n \rightarrow \{ 0,1 \}$ is \emph{monotone} if 
$${\bf x} \leq {\bf y} \Longrightarrow f({\bf x}) \leq f({\bf y}).$$

By identifying $\{ 0, 1\} ^n$ with $\wp (\{ 1,\ldots ,n \} )$, the
relation $\leq$ on $\{ 0,1 \} ^n$ corresponds to inclusion
($\subseteq$) in $\wp ( \{ 1,\ldots ,n \} )$. Then a monotone function $f$
corresponds to a function from $\wp (\{ 1,\ldots ,n \} )$ to $\{ 0,1 \}$
such that $A \subseteq B \Longrightarrow f(A) \leq f(B)$.

A monotone function $f$ naturally defines an adversary structure
${\cal A}_f = \{ B \subseteq P: f(B)=0 \}$.

Given an adversary structure ${\cal A}$ and a monotone function $f$,
we say $f$ \emph{rejects} ${\cal A}$ if $f(B)=0$ for all $B \in {\cal
  A}$, that is if ${\cal A} \subseteq {\cal A}_f$.

With these definitions in hand we can state the complexity of the 
Hirt-Maurer protocols: their generalization of \cite{RB89} runs in time 
$m^{O(\log \log m)}$, where $m$ is the size of the smallest monotone 
formula consisting of majority-accepting gates which rejects the adversary structure ${\cal A}$.

\subsection{Monotone span programs}
\label{sec:MSP}

Span programs were introduced as a model of computation in
\cite{KW93}. They were first used as a tool for multiparty computation
by Cramer, Damg{\aa}rd and Maurer \cite{CDM98}. In this section we define the concepts related to
monotone span programs relevant to this paper.

\begin{definition}
  A {\bf {monotone span program}} (MSP) over a set $P$ is a triple
  $(K,M,\psi)$ where $K$ is a finite field, $M$ is a $d \times e$
  matrix over $K$ and $\psi: \{1, \ldots, d \} \rightarrow P$ is a 
  surjective function.
\end{definition}

The MSP associates to each subset $B \subseteq P$ a subset of the rows
of $M$: the set of rows $l$ such that $\psi(l) \in B$. This
corresponds to a linear subspace of $K^e$ (the span of those rows).
The monotone function $f: \wp (P) \rightarrow \{ 0,1\}$ defined by a
MSP is given by the rule $f(B)=1$ if and only if the ``target vector''
$\epsilon = (1,0,0,\ldots,0)$ is in the subspace associated with $B$.
If we denote by $M_B$ the submatrix of $M$ formed of the rows $l$ such
that $\psi (l) \in B$ then we get that
$$f(B) = 1 \iff \epsilon \in Im(M_B^T).$$

Given a MSP computing $f$, there is a secret sharing scheme which
tolerates the corresponding adversary structure ${\cal A}_f$. This
scheme is explained in section \ref{sec:SS}. 

The definition above is sufficient for ``secret sharing''-type
protocols such as VSS and for multiparty computations in which
multiplication in the field is not necessary. For general MPC,
however, we need a stronger notion.

\begin{definition}[(due to \cite{CDM98})] 
  A MSP $(K, M, \psi)$ is said to be {\bf with multiplication} if
  there exists a vector ${\bf r}$ (called a ``recombination vector'')
  such that
  $$\forall {\bf b}, {\bf b'} \in K^e : \quad \langle {\bf r}, M {\bf
    b}*M {\bf b'} \rangle = \langle \epsilon,{\bf b}*{\bf b'}
  \rangle$$
  where $\epsilon = (1,0,\ldots,0)$, $\langle \cdot , \cdot
  \rangle$ is the standard inner product on $K^e$ and for ${\bf v} =
  (v_1,...,v_d), {\bf w} = (w_1,...,w_d),$ we have ${\bf v}*{\bf w}
  =(v_1w_1,...,v_dw_d)$.
\end{definition}

In \cite{CDM98} it is proved that for any ${\cal Q}^2$ adversary
structure ${\cal A}$, one can construct a MSP with multiplication
which rejects ${\cal A}$. The MSP can be constructed so it is linear
in the size of the smallest majority-accepting formula rejecting ${\cal A}$.

Note that a counting argument shows that not all families of ${\cal
  Q}^2$ adversary structures over $n$ players (for $n=1,2,\ldots$) can
be rejected by a family of MSP's with size polynomial in $n$.

See the open questions in section \ref{sec:questions} for further
discussion.

\subsection{Previous work}
\label{sec:cdm}

This work follows the initiative of Cramer, Damg\r ard and Maurer in
\cite{CDM98} for adapting existing threshold protocols to general
adversary structures using monotone span programs. In that paper the
results of \cite{GMW87} and \cite{BGW88,CCD88} were adapted to ${\cal
  Q}^2$ and ${\cal Q}^3$ structures respectively. We state their genralization of 
\cite{BGW88,CCD88}: 

\begin{theorem}
  \label{thm:Q3}
  Let ${\cal A}$ be a ${\cal Q}^3$ adversary structure and $\pi$ some
  multi-party protocol agreed upon $n$ players. Let $(K,M,\psi)$ be a
  MSP \emph{with multiplication} rejecting ${\cal A}$ and suppose
  $\pi$ can be implemented in $s$ steps with operations over $K$.

  Then there is a protocol for $\pi$ tolerating ${\cal A}$ which is
  information-theoretically secure and which has complexity polynomial
  in $log | K |, s$ and the size of $M$.
\end{theorem}

\subsection{This article}
\label{sec:report}

In this paper we adapt the results of \cite{RB89} to ${\cal Q}^2$ 
structures with information-theoretic security.
As mentioned above, this had already been done by Hirt and Maurer in
\cite{HM97} without using MSP's. However, their protocol ran in time
$m^{O( \log \log m)}$ where $m$ is the size of the smallest monotone
formula consisting of majority-accepting gates which rejects the adversary
structure ${\cal A}$.

Our protocol on the other hand is polynomial in the size of the
smallest MSP with multiplication rejecting ${\cal
  A}$. Since MSP's with multiplication are at least as efficient as
majority-accepting formulae (proved in \cite{CDM98}), our protocol is more efficient than the \cite{HM97} construction.

The main results are:

\begin{theorem}
  \label{thm:VSS}
  Let ${\cal A}$ be a ${\cal Q}^2$ adversary structure on $n$ players
  and $(K,M,\psi)$ be a MSP rejecting ${\cal A}$. Suppose a reliable
  broadcast channel and secure communication between every pair of
  players is available.
  
  There is a VSS scheme for $n$ players, tolerating ${\cal A}$, which
  has error probability $\leq 2^{-k}$ and which has complexity
  polynomial in $log | K |, n, k$ and the size of $M$.
\end{theorem}

\begin{theorem}
  \label{thm:q2MPC}
  Let ${\cal A}$ be a ${\cal Q}^2$ adversary structure and $\pi$ some
  multi-party protocol agreed upon $n$ players. Let $(K,M,\psi)$ be a
  MSP \emph{with multiplication} rejecting ${\cal A}$ and suppose
  $\pi$ can be implemented in $s$ steps with operations over $K$.
  Suppose a reliable broadcast channel and secure communication
  between every pair of players is available.
  
  Then there is a protocol for $\pi$ tolerating ${\cal A}$ which has
  error probability $\leq 2^{-k}$ and which has complexity polynomial
  in $log | K |, s, n, k$ and the size of $M$.
\end{theorem}

Our protocol follows the construction of \cite{RB89}. We first present
the basic secret sharing scheme using $MSP$ as well as an information 
checking
protocol. We then give a protocol for \emph{weak secret sharing} which
we use to build a protocol for \emph{verifiable secret sharing}. Using
these tools we present the protocol for general MPC. For our protocol
to be efficient, we change the product checking protocol of
\cite{RB89} \footnote{The protocol given here was published in a
  different context (computational proofs of knowledge) in
  \cite{ChaumEG1987}. We independently ``discovered'' a slightly
  different version in 1998.}.  The protocol we give is polynomial in
$\log |K|$ whereas that of \cite{RB89} is $\Omega(|K|^2)$. We conclude
with some open questions.

\section{Secret Sharing and Information Checking}
\label{sec:SSIC}

\subsection{Secret Sharing}
\label{sec:SS}

Given a MSP $(K,M,\psi)$, we can define a secret sharing scheme which
tolerates the adversary structure ${\cal A}_f$ induced by the MSP (see
section \ref{sec:MSP}).  Recall that $M$ is a $d \times e$ matrix over
the field $K$ and $\psi: \{ 1,\ldots , d\} \rightarrow \{
1,\ldots,n\}$ is an arbitrary function.

Say the dealer has a secret $a \in K$. He extends it to an $e$-rowed
vector by adding random field elements $\rho_2,\ldots,\rho_e$ to make
a vector ${\bf a_*} = (a,\rho_2,\ldots,\rho_e)$. Let {\bf$\alpha$} $=
M {\bf a_*}$ and let ${\bf \alpha}_A$ denote the elements of $\alpha$
with indices in $A$ where $A \subseteq \{ 1,\ldots , d\}$. Then the
dealer gives $\alpha_l$ to player $P_{\psi(l)}$. In the end, each
$P_i$ receives ${\bf \alpha}_{\psi^{-1}(i)}$.

From now on this protocol will be referred to as SHARE$(D,a)$ where
$D$ is the dealer.

\begin{lemma}
  SHARE is a secret sharing scheme secure against ${\cal A}_f$. That
  is, no coalition in ${\cal A}_f$ can learn any information about the
  secret but any set of players \emph{not} in ${\cal A}_f$ can
  reconstruct it.
\end{lemma}

\begin{proof}
  See \cite{CDM98}.
\end{proof}

\subsection{Information Checking}
\label{sec:IC}

The protocol in this section is based on \cite{RB89}. All computations
are done over $F = GF(3^k)$ where $k$ is the security parameter. We
require that $3^k > |K|^d$. This allows the encoding of any set of
shares from the secret scharing scheme induced by the MSP
$(K,M,\psi)$.

We will use in the sequel a Guaranteed Information Checking (GIC)
protocol:
\begin{description}
\item[Pre:] $D$ has already sent $INT$ his secret $s \in F$.
  
\item[Post:] $INT$ is guaranteed that an honest $R$ ($D$ may be
  dishonest) will always (i.e with very high probability) accept his
  value for $s$. Moreover, no information about $s$ is leaked as long
  as $D$ and $INT$ are honest.
  
\item[Protocol:] GIC-Generate$(D \rightarrow INT \rightarrow R,s)$
  \begin{enumerate}
  \item $D$ makes $2k$ vectors $(y_i, b_i,c_i)$ such that $b_i \in_R
    F-\{0\}, y_i \in_R F$ and $c_i = s + b_i y_i$. He sends $s$ and
    the $y_i$ to $INT$ and sends the check vectors $(b_i,c_i)$ to $R$.
  \item $INT$ picks a random set $I \subseteq \{ 1,\dots,2k \}$ such
    that $|I| = k$ and broadcasts $I$.
  \item $R$ broadcasts the check vectors $(b_i,c_i)$ with $i \in I$.
  \item \label{Dcheck} $D$ checks whether or not this is indeed what
    he sent to $R$. If so, he broadcasts his approval. If not, he
    creates a new triple $(y,b,c)$ such that $c = s + by$ and $b \neq
    0$.  He sends $y$ to $INT$ and \emph{broadcasts} the single check
    vector $(b,c)$.
  \item Based on what he has seen, $INT$ ``guesses'' whether or not
    $R$ will now accept his value. If $D$ approved in the previous
    step then $INT$ decides ``YES'' if and only if $R$'s pairs all
    agreed with the values $INT$ possesses. If $D$ disapproved and
    created a new check vector, $INT$ outputs ``YES'' if and only if
    $c = s + by$ actually holds.
  \item If $INT$ thinks his value will be accepted by (an honest) $R$
    he broadcasts his approval. If not, $INT$ asks $D$ to broadcast
    $s$.
  \end{enumerate}
  
\item[Protocol:] GIC-Authenticate$(INT \rightarrow R, s)$.
  \begin{enumerate}
  \item $INT$ sends $s$ along with either $\{ y_i: i \not \in I \}$ or
    $y$ (depending on what occurred at step \ref{Dcheck}) to $R$.
  \item $R$ accepts if any one of the $y_i$'s agrees with her
    corresponding pair $(b_i,c_i)$, or if $y$ agrees with $(b,c)$.
  \end{enumerate}
\end{description}

Notice that at the end of GIC-Generate, $INT$ is guaranteed (with high
probability) that an honest $R$ will accept his value should he send
it to $R$ later on. Also notice that if $D$ and $INT$ are honest, no
other party will gain any information about $s$ (including $R$).

\begin{lemma}
  The GIC protocols have error probability less than $2^{-k}$.
\end{lemma}

\begin{proof}
  See \cite{RB89}.
\end{proof}

\section{Weak Secret Sharing}
\label{sec:WSS}

This WSS scheme comes (essentially) from \cite{RB89}.

Before describing the protocol note that we will refer to the WSS of a
value $a$, by $D$, as $[a]^W_D$. A similar VSSed value will be denoted
$[a]^V_D$ and a verifiably shared secret belonging to no particular
player will be written $[a]^V$.

From now on we will always assume that the MSP being used is ${\cal
  Q}^2$, that is we assume that the adversary structure ${\cal A}_f$
induced by the MSP is ${\cal Q}^2$.

The WSS scheme is in two parts: the commitment protocol (WSS) and the opening protocol (WSS-OPEN).

\begin{description}
\item[Pre:] The dealer $D$ has a secret $a \in K$.
  
\item[Post:] $D$ has shared $a$ such that either
  \begin{itemize}
  \item The shares the honest players hold are consistent with a
    single value which $D$ can reveal, or
  \item The shares are inconsistent and $D$ will always be caught and
    disqualified during the WSS-OPEN protocol.
  \end{itemize}
  
  Moreover, an honest $D$ will never be disqualified.
  
\item[Protocol:] WSS$(D,a)$
  \begin{enumerate}
  \item SHARE$(D,a)$
  \item For every $i \neq j$: GIC-Generate$(D \rightarrow P_i \rightarrow P_j,
    \alpha_{\psi^{-1}(i)} )$.
  \end{enumerate}
  
  Notice this protocol guarantees $P_i$ that at some later time he can
  transmit his share to $P_j$ and she will be convinced that $D$
  indeed gave him ${\bf \alpha}_{\psi^{-1}(i)}$.
\end{description}

Based on this protocol we can define a \emph{weakly shared value} to be a
value $a$ which a dealer $D$ has shared (not necessarily correctly)
such that GIC has been run for every pair $(P_i,P_j)$ with $i \neq j$.

We now give the opening protocol.

\begin{description}
\item[Pre:] $a$ is weakly shared by $D$.
\item[Post:] There is a single value $a$ which $D$ can reveal. All the
  honest players will output the same value, which will be either $a$
  or $null$. They will output $null$ only if $D$ has acted dishonestly
  in sharing or revealing the secret.
\item[Protocol:] WSS-OPEN$(D,[a]^W_D)$
  \begin{enumerate}
  \item $D$ broadcasts the vector ${\bf a_*}$ he created during the
    SHARE protocol.
  \item Each $P_i$ runs GIC-Authenticate with $P_j$ \\
    (Thus $P_j$ obtains ${\bf \alpha}_{\psi^{-1}(i)}$ if $P_i$ is
    honest and rejects the value if $P_i$ tries to cheat.  In the end,
    an honest $P_i$ will have obtained $\alpha_{\theta}$ where
    $f(\theta) = 1$, so he can reconstruct the secret if the shares of
    honest players are consistant.)
  \item If for any $i$ such that $P_j$ accepted $P_i$'s value there is
    an inconsistency (i.e. ${\bf \alpha}_{\psi^{-1}(i)} \neq
    M_{\psi^{-1}(i)} {\bf a_*}$) then $P_j$ accuses $D$.
  \item If the set of accusers is not in the adversary structure (i.e.
    $f(\{accusers\}) =1$) then $D$ is disqualified.  Otherwise his
    value (that is the first coordinate of ${\bf a_*}$) is accepted.
  \end{enumerate}
\end{description}

Notice that $D$'s cooperation is essential to opening the commitment
and that if need be, $D$ can open his WSS only to a single player
$P_i$ by having all players send information only to $P_i$.

\begin{lemma}
  For any MSP whose adversary structure ${\cal A}_f$ is ${\cal Q}^2$,
  the pair of protocols (WSS, WSS-OPEN) is an ${\cal A}_f$-secure weak
  secret sharing scheme with error probability exponentially small in
  $k$.
\end{lemma}

\subsection{Linear operations on weakly shared values}
\label{sec:WSS-add}

It is clear that a WSSed value can be multiplied by any constant
$\lambda$: each INT multiplies his share by $\lambda$ and each $R$
multiplies each of his pairs $(b,c)$ by $\lambda$. Denote such a
multiplication by $[\lambda a]^W_D \leftarrow \lambda * [a]^W_D$.

To add two WSSed values belonging to the same dealer, each player adds
his shares of the two secrets to obtain his share of their sum. Then
do GIC-Generate$(D \rightarrow P_i \rightarrow P_j,{\bf
  \gamma}_{\psi^{-1}(i)} )$ where {\bf $\gamma$} is the vector of
shares of the sum. The result is a WSS of the sum. Denote this by
$[a+b]^W_D \leftarrow [a]^W_D + [b]^W_D$.

\begin{remark}
  If $D$ does not commit $a$ properly but does commit $b$ properly, he
  will be caught as a cheater if he opens $a+b$. This yields a simple
  zero-knowledge proof that a value is correctly committed by WSS.
  Have $D$ pick $b$ at random and commit to $b$ and then use the
  preceding protocol to obtain $[a+b]^W_D$. Then flip a coin and have
  him either open $b$ or $a+b$ depending on the outcome.  If he was
  badly comitted to $a$ he will be caught with probability 1/2. Repeat
  the protocol $k$ times to get exponentially small probability of
  error.
\end{remark}

\section{Verifiable Secret Sharing}
\label{sec:VSS}

Verifiable secret sharing is a primitive introduced in
\cite{ChorGMA85}. The scheme we give comes essentially from
\cite{RB89}.

\begin{description}
\item[Pre:] The dealer $D$ has a secret $a \in K$.

\item[Post:] $D$ is committed to a unique value which can be
  efficiently recovered without his help. Moreover, each player has
  committed to his share by means of a WSS. The shares of all players
  at the top (VSS) level are consistent. The shares of honest players
  at the lower (WSS) level are consistent.
  
\item[Protocol:] VSS$(D,a)$

  \begin{enumerate}
  \item SHARE$(D,a)$.
    
  \item For $l=1,..,d$ do WSS$(P_{\psi(l)} , \alpha_l)$.
    
  \item For $j=1,..,kn$ do:
    \begin{enumerate}
    \item $D$ chooses $c^{(j)} \in_R K$
    \item SHARE$(D, c^{(j)} )$ (yielding a random vector ${\bf
        c}^{(j)}_*$ and shares $\gamma^{(j)}_1,\ldots,\gamma^{(j)}_d$)
    \item For $l=1,..,d$ do:
      \begin{itemize}
      \item WSS$(P_{\psi(l)} , \gamma^{(j)}_l)$
      \item $[\gamma^{(j)}_l + \alpha_l]^W_{P_{\psi(l)}} \leftarrow
        [\gamma^{(j)}_l]^W_{P_{\psi(l)}} + [\alpha_l]^W_{P_{\psi(l)}}$
      \end{itemize}
    \end{enumerate}
    
  \item For $j=1,..,kn$ do:
    \begin{enumerate}
    \item $P_{j\textrm{ mod } n}$ flips a coin and broadcasts the
      result.
    \item
      \begin{description}
      \item[Heads:] Let
        \begin{itemize}
        \item $b = c^{(j)}$
        \item ${\bf b_*} = {\bf c}^{(j)}_*$
        \item $\beta = \gamma^{(j)}$
        \end{itemize}
      \item[Tails:] Let
        \begin{itemize}
        \item $b = a + c^{(j)}$
        \item ${\bf b_*} = {\bf a}_* + {\bf c}^{(j)}_*$
        \item $\beta = \alpha + \gamma^{(j)}$
        \end{itemize}
      \end{description}
    \item $D$ broadcasts ${\bf b_*}$.
    \item Each $P_i$ checks if $\beta_{\psi^{-1}(i)} =
      M_{\psi^{-1}(i)}{\bf b_*}$. If not, $P_i$ accuses $D$. $D$ must
      then broadcast all information given to $P_i$, that is
      $\alpha_{\psi^{-1}(i)}$ and $\gamma^{(j)}_{\psi^{-1}(i)}$ for
      all $j$.  $P_i$ is removed from the VSS protocol (if $D$ does
      not broadcast the requested information, he is deemed corrupt).
    \item For $l=1,..,d$ do WSS-OPEN$(P_{\psi(l)},[\beta_l]^W)$ (as
      long as $P_{\psi(l)}$ remains in the protocol).

      If $P_{\psi(l)}$ gets caught in WSS-OPEN or if the value he
      reveals is inconsistent with the ${\bf b}_*$ broadcasted by $D$ then
      he is deemed corrupt and is removed from the protocol. All his
      shares of $a$ and of all the $c^{(j)}$ are then broadcasted by
      $D$.
    \end{enumerate}
    
  \item If the set of participants who are removed from the protocol
    (at any step) is qualified (i.e. not in the adversary structure)
    or if $D$ broadcasts inconsistent information then $D$ is deemed
    corrupt and the VSS is considered failed.  Otherwise the VSS is
    deemed a success.
  \end{enumerate}
  
  Note that as long as no errors occur in the subprotocols, the only
  way for $D$ to succeed in passing off inconsistent shares for $a$ is
  to correctly guess all the coin flips. Since at least one player is
  honest at least $k$ of the coin flips are fair and so the failure
  probability is below $2^{-k}$.

\end{description}
  
Based on this protocol we can define a \emph{verifiably shared value}
to be a value $a$ such that every player is committed to his share of
$a$ via WSS. Moreover, the shares of all players at the top level must
be consistent as must the shares of honest players at the bottom (WSS)
level.

The opening protocol is VSS-OPEN:

\begin{description}

\item[Pre:] $a$ is a verifiably shared value.
\item[Post:] All honest players output $a$.

\item[Protocol:] VSS-OPEN$([a]^V)$

  \begin{enumerate}
  \item Each player opens his WSS to his share of the secret.
  \item $a$ is reconstructed from any qualified set of players who
    opened their WSS succesfully or whose shares had been broadcast in
    the VSS protocol.
  \end{enumerate}
  
  Notice that no false shares can be contributed since any bad WSS's
  would have been detected (with high probability) in the VSS
  protocol. Moreover, all honest players \emph{will} reveal their
  secret correctly and so a qualified set of shares \emph{is}
  available for reconstruction.
  
  Also notice that $D$'s participation is not necessary and that the
  OPEN protocol works for any verifiably shared secret.
\end{description}

\begin{lemma}
  For any MSP whose adversary structure ${\cal A}_f$ is ${\cal Q}^2$,
  the pair of protocols (VSS, VSS-OPEN) forms an ${\cal A}_f$-secure
  VSS scheme with error probability exponentially small in $k$.
\end{lemma}

This proves theorem \ref{thm:VSS}.

\subsection{Linear operations on verifiably shared values}
\label{sec:VSS-ADD}
      
It is possible to perform linear operations on a verifiably shared
value by performing the corresponding operations on the shares
(committed to via a WSS). In the case of VSS, it is \emph{not}
necssary that the secrets being added belong to the same dealer or
indeed to anyone at all.

\subsection{Converting WSS to VSS}
\label{sec:WSS2VSS}

A shared value $[a]^W_D$ can be converted to $[a]^V_D$ by throwing
away the check vector information of the GIC protocols and considering
the WSS as a simple secret sharing. The VSS protocol can then be
started from step 2. The cooperation of the dealer \emph{is} required
for converting his WSS to a VSS.

\section{Multi-Party Computation}
\label{sec:MPC}

The protocol for computing a function $g(x_1,x_2,\ldots , x_n)$ where
$x_i$ is the input of $P_i$ follows the basic outline of
\cite{BGW88,CCD88,RB89}. Before the computation begins, the players
decide on an arithmetic circuit over $K$ which computes $g$. Each
player commits to his input via a VSS $[x_i]^V_{P_i}
$. The players then
evaluate the circuit gate by gate to eventually end up with
$[g(x_1,x_2,\ldots ,x_n)]^V$. This commitment is then opened publicly.

We already discussed how to achieve a multi-party computation for
addition and multiplication by a constant, so all that is missing now
is a multi-party multiplication protocol.

\subsection{Checking a product}
\label{sec:VSS-CP}

We start by describing a protocol VSS-CP used by a dealer to prove
that three secrets $[a]^V_D, [b]^V_D$ and $[c]^V_D$ satisfy $ab = c$.
This protocol replaces the \cite{RB89} protocol that used the
multiplication table of the field $K$. The \cite{RB89} protocol is
insufficient since it runs in time $\Omega(|K|^2)$ whereas we require
a protocol polynomial in $\log |K|$. We require this since it may be
that the only polynomial sized MSP's for a given adversary structure
happen to be over large fields (see open questions in section
\ref{sec:questions} for further discussion).

The protocol given here appeared in \cite{ChaumEG1987,BoyarCDP1990},
for commitments based on the discrete logarithm problem. As it
appears here it works for any homomorphic commitment scheme (i.e. one
which allows addition of secrets).

\begin{description}
\item[Pre:] The dealer has $[a]^V_D$, $[b]^V_D$ and $[c]^V_D$ where
  $ab=c$.
\item[Post:] Every participant, knowing only shares of $[a]^V_D$,
  $[b]^V_D$ and $[c]^V_D$, will be convinced (with very small
  probability of error) that $ab = c$.
  
\item[Protocol:] VSS-CP$(D, [a]^V_D,[b]^V_D,[c]^V_D)$

  \begin{enumerate}
  \item Repeat for j = 1,...,kn,
    \begin{enumerate}
    \item D chooses $b'$ $\in_{R} K$ and computes $c' = ab'$.
    \item D commits himself to $b'$ and $c'$ by computing
      \begin{itemize}
      \item $[b']^V_D \leftarrow$ VSS$(D,b')$
      \item $[c']^V_D \leftarrow$ VSS$(D,c')$
      \end{itemize}
    \item The participant $P_{j{\bf mod}n}$ flips a coin:
      
      \begin{enumerate}
      \item If {\bf Heads}:
        \label{step:CPheads}
        \begin{itemize}
        \item The participants open $[b']^V_D$.
        \item They collectively compute $[ab' - c']^V_D \leftarrow
          b'*[a]^V_D - [c']^V_D$.
        \item They open this commitment and check it is 0.
        \end{itemize}
      \item If {\bf Tails}:
        \label{step:CPtails}
        \begin{itemize}
        \item The participants collectively compute and open $[b +
          b']^V_D$.
        \item They collectively compute $[a(b + b') - (c + c')]^V_D
          \leftarrow (b + b')*[a]^V_D - [c']^V_D - [c]^V_D$.
        \item They open this commitment and check it is 0.
        \end{itemize}
      \end{enumerate}
    \end{enumerate}
  \end{enumerate}
  
\item[Analysis:] If in fact $c = ab$, the verification in steps
  \emph{\ref{step:CPheads}} and \emph{\ref{step:CPtails}} will always
  be successful.  On the other hand, if $c \neq ab$, there are two
  cases. First, if in fact $D$ chose $c' = ab'$, in step
  \ref{step:CPtails} the participants would find that $c + c' \neq ab
  + ab'$.  Second, if $c \neq ab$ but $c' \neq ab'$, the verification
  in step \ref{step:CPheads} would fail. The protocol VSS-CP thus
  provides a proof to each honest player with probability of error
  less then $2^{-k}$. Moreover this proof is zero-knowledge in that
  the information revealed is never enough to reveal any information
  about $c$, $a$ or $b$.

\end{description}

\subsection{The Multiplication protocol}
\label{sec:MULT}

This protocol is the multiplication protocol for active adversaries in
\cite{CDM98}. It assumes that the MSP being used has multiplication
(see section \ref{sec:MSP} for details).

\begin{description}
  
\item[Pre: ] We have $[u]^V, [v]^V$.
\item[Post:] We obtain $[uv]^V$, that is $uv$ is a verifiably shared
  secret.
  
\item[Protocol:]MULT$([u]^V, [v]^V)$ We will denote by ${\bf \mu}$ and
  ${\bf \nu}$ the vectors that share $u$ and $v$ respectively.
\begin{enumerate}
\item For $l=1,..,d$ do
        \begin{itemize}
        \item $[\mu_l]^V_{\psi(l)} \leftarrow [\mu_l]^W_{\psi(l)}$
        \item $[\nu_l]^V_{\psi(l)} \leftarrow [\nu_l]^W_{\psi(l)}$
          
          (see section \ref{sec:WSS2VSS} for how to do this).
        \end{itemize}   
      \item For $l=1,...,d$, do:
        \begin{itemize} 
        \item Player $P_{\psi(l)}$ computes $\omega_l := \mu_l\nu_l$
        \item $[\omega_l]^V_{\psi(l)} \leftarrow \textrm{VSS}
          (P_{\psi(l)},\omega_l)$
        \item Run VSS-CP$(P_{\psi(l)},[\mu_l]^V_{\psi(i)},
          [\nu_l]^V_{\psi(i)}, [\omega_l]^V_{\psi(l)})$,\\
          so as to prove to everybody that his commited value
          $\omega_l$ is in fact $\mu_l\nu_l$.
        \end{itemize}
      \item For $l=1,..,d$ do
        \begin{itemize}
        \item Collectively compute
          $$[uv]^V = r_1 * [\omega_1]^V_{\psi(1)} + \cdots + r_d *
          [\omega_d]^V_{\psi(d)},$$
          where ${\bf r}=(r_1,...,r_d)$ is
          the recombination vector (this is a simple linear
          combination and can be done as in section
          \ref{sec:VSS-ADD}).
        \end{itemize}     

\end{enumerate}

If at any stage of the computation, a participant is detected as
beeing a cheater, he is excluded from the protocol.  The only problem
that may arise is in step 1, since the participation of the dealer is
necessary to convert a WSS to a VSS. If this ever happens, we simply
reset the protocol to the input distribution stage, the remaining
players open the cheaters' VSSed inputs and then they restart the
protocol.  This will prolong the protocol by a factor of at most $n$.

Since $uv$ is now a verifiably shared secret, it can be efficiently
opened by the honest players using VSS-OPEN.

Notice the protocol given has no inherent probability of error. Its
probability of error is at most the sum of the error probabilities of
its subprotocols. As each of these has error probability exponentially
small in $k$ and is executed polynomially many times, the error
probability remains exponentially small in $k$.
\end{description}

This completes the proof of theorem \ref{thm:q2MPC}.

\section{Open questions}
\label{sec:questions}

\begin{enumerate}

\item It has not yet been proven whether MSP's \emph{with
multiplication} are super-polynomially more efficient than majority
accepting circuits (or formulae) for computing max-${\cal Q}^2$
functions\footnote{A max-${\cal Q}^2$ adversary structure is one to
which no more sets can be added without it losing the ${\cal Q}^2$
property. A max-${\cal Q}^2$ function is one whose associated
adversary structure ${\cal A}_f$ is max-${\cal Q}^2$}. In
\cite{CDM98}, a super-polynomial gap has been proved for general MSP's
(which don't necessarily have the multiplication property). An
interesting question is to determine how MSP's with multiplication
perform as compared to MSP's without multiplication.
  
  One thing which is known is that schemes based on MSP's with
  multiplication (or even with strong multiplication) are at least as
  efficient as those based on threshold formulae. In particular this
  implies that the protocol given here is at least as efficient as the
  one in \cite{HM97}. This is proved in \cite{CDM98}.
   
\item No results have so far been published which extend threshold
  results other than \cite{GMW87}, \cite{BGW88}, \cite{CCD88} and
  \cite{RB89}. Although it would seem that the ideas from \cite{CDM98}
  extend more or less directly to many distributed threshold
  protocols, this is not the case for asynchronous protocols.
 
  In asynchronous systems, MPC is possible tolerating any active
  ${\cal Q}^3$ adversary structure (by extending results of
  \cite{Bra87,BKR1994}). However, no polynomial-time protocol
  currently exists. The bottleneck is a primitive from distributed
  computing known as Byzantine Agreement (BA). Although an efficient
  asynchronous BA protocol exists for threshold structures with $t <
  n/3$ (from \cite{CR93}), the proof of correctness given there does
  not carry over to general structures.

\item Very few results exist which connect the size of the field used
in an MSP to the size of the matrix $M$. Although one can pass from
$K$ to a subfield with blowup only quadratic in the degree of the
extension \cite{Cramer98}, it is possible that certain functions have
polynomial size MSP's which work only over unmanageably large prime
fields.

\end{enumerate}

\section*{Acknowledgements}
\label{sec:ack}

We would like to thank Claude Cr\'epeau for his support and many
helpful comments, as well as for suggesting this area of study, as
well as Ronald Cramer and Ivan Damg{\aa}rd for their comments on the
manuscript.

\nocite{KW93}

\bibliographystyle{alpha}
\bibliography{crypto,monotone}

\end{document}